\documentclass[11pt,a4paper]{article}

\usepackage{graphicx} 
\usepackage{float} 
\usepackage{amssymb}
 \usepackage{amsmath}
 \usepackage{amsthm}
 
\newtheorem*{defi}{Definition}{\bf}{\it}
\newtheorem*{obs}{Observation}{\bf}{\it}
{\bf}{\it}
{\bf}{\it}

\graphicspath{{PICTS/}}

\title{Computing with Cognitive States}
 
\author{
Stefan Reimann  \\
  Department of Psychology - Cognitive Psychology,\\
  University of Zurich,\\
  ~\\ 
  Institute of Neuroinformatics\thanks{email: reimannst@ini.uzh.ch}\\
University of Zurich and ETH Zurich
  }

\begin{document}

\maketitle

\begin{abstract}
Basic experimental findings about human working memory  can be described by an algebra built on high-dimensional binary states, representing information items, and two operations: multiplication for binding and addition for bundling. In contrast to common VSA algebras, bundling is not associative. Consequently bundling a sequence of items preserves their sequential ordering. The cognitive states representing a memorised list exhibit a primacy as well as a recency gradient. The typical concave-up and asymmetrically shaped serial position curve is derived as a linear combination of those gradients. Quantitative implications of the algebra are shown to agree well with empirical data from basic cognitive tasks including storage and retrieval of information in human working memory. 

\textbf{Keywords:} 
human working memory; activation gradients; serial position curve; holographic representation; high-dimensional computing
\end{abstract}

\section{Introduction}
Cognitive functions are established by the interplay of processes in the brain, in which patterns of neuronal activity interact and are consecutively transformed. They can be excited by some input and might finally evoke or suppress some behavioural output. Related computations are governed by simple binary units, e.g. neurons, which constitute a highly complex medium, i.e. a huge heterogeneous and irregular synaptic network. Thus, complexity is in the medium rather than in the units. Transferring this view to the question about how an algebra governing these computations should look, leads to the framework of high-dimensional computing: The algebra governing such computation is defined on a complex, i.e. high dimensional and random space, while rules for bundling or collecting information and multiplication for binding information items together are elementary binary operations. 

Experimental research has revealed a number of typical findings about the functioning of memory. According to the above view, the 'cognitive algebra;' to be proposed should be able to reproduce such findings. The question therefore is: "Can the algebra reproduce the experimental data about memory, and how far can be go with this algebra alone." Thus the aim is to analyse the corresponding algebra alone, especially with regard to its congruence with empirical findings including the serial position curve. 

Human Working Memory is commonly regarded as a functional subsystem of memory, whose goal is to hold and to organise information for some short period of time in order to make it available for higher cognitive processes \cite{cowan2017many}. Experiments in this field rely on the subtle construction of input data such as memory lists and produce output data such as recall probabilities or response times \cite{murdock1974human,kahana2012foundations,oberauer2018benchmarks}. 
Among these, the most prominent finding is the serial-position curve, which shows the accuracy of item retrieval varying as a function of serial position in a memory list, averaged over a sample of participants. As observed across (probably all) immediate memory tasks, it has a concave-up shape and is asymmetric. Its particular shape depends on the particular cognitive task. such as recognition, free recall, backward and forward serial recall, or cued recall. 
 For example in recognition and in cued (probed) recall the serial position curve shows a strong recency effect, while the primacy effect is weak. Strong primacy effects are seen in forward recall, while recency effects are strong in backward or in free recall. 

\begin{figure}[h]\label{fig:expCurve}
{\center
		\includegraphics[height=3in]{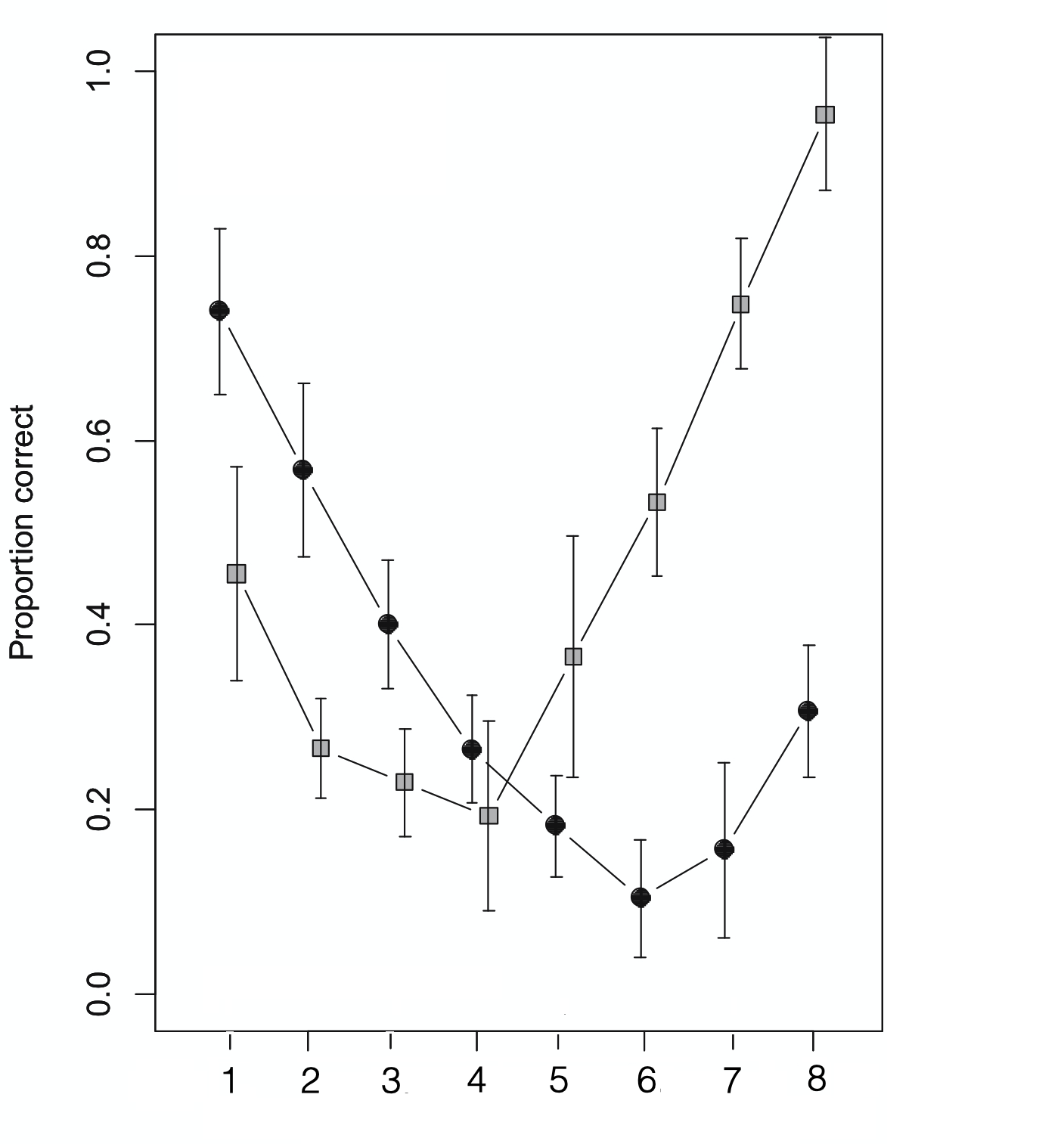} 
		\caption{{\bf Primacy effect and recency effects}:  
		Data are for immediate forward and backward serial recall %
		but are similarly in other immediate  tasks.}
		}
\end{figure}

To describe particular aspects of the functioning of the human working memory, models with different characteristics have been used, differing both in terms of the medium in which the information is stored and the storage operations used. Models include local code models such as REM \cite{shiffrin1997model}, distributed models of memory such as SOB \cite{farrell2002endogenous} and TCM \cite{howard2002distributed}, as well as holographic models such as TODAM \cite{murdock1982theory,murdock1993todam2} which uses high-dimensional probabilistic encoding for the holographic representation of information \cite{plate1991holographic}. Holographic models gain from the  properties, which are implied by high dimensionality together with randomness, see \cite{kanerva2009hyperdimensional} for an overview about the framework of high-dimensional computing. 

The holographic approach appears as a natural candidate to model the functioning of cognitive processes. Input items evoke activity patterns in the respective neural field; 
The fact that these representations are sparse and the consequences thereof are not explicitly considered in this note.
Computation consists in transforming those patterns according to two elementary operations: The additive-like superposition realises the bundling of item information, while multiplication realises binding of items. The high-dimensional space of binary patterns together with these two operations form a high-dimensional algebra governing storage and computation in this system. 

Before giving an outline of the paper, a remark seems worthwhile: The aim is not to provide a full-blown model rather than to propose an elementary computational structure, an algebra, on top of which a model could be constructed. The main question is, how much of experimental findings can already be described on the basis of that algebra alone. 

The outline of this paper is briefly as follows: Firstly, the state-space is defined as a high-dimensional Hamming space (eq \ref{eq:state-space}) equipped with some distance on it. A similarity measure is proposed which is derived from that distance. It allows both, to judge about the familiarity of two states as well as about their distinctiveness (eq \ref{eq:similarity}). Computing is by manipulating states according to two operations on that space: multiplication for binding and a not associative addition for bundling. This completes the definition of the algebra (eq \ref{eq:algebra}) to be considered. 
Non-associativity is an essential feature of that bundling since it implies that the sum of components depends on their sequential ordering (eq \ref{eq:non-asso}). As a consequence, information about the order of sequentially presented list items is conserved. 
The corresponding left-associative sum and the right- associative sum of list items correspond to states exhibiting a recency and a primacy gradient, respectively (Fig \ref{fig:ActivationGradients}). As applications basic cognitive tasks such as item recognition and probed recall are considered. The typical concave-up and asymmetrical shape of the serial-position curve is derived as a mixture of these two activation gradients (Fig \ref{fig:SPC}).

\section{The algebra of cognitive states $\left( \mathbb{X}, +_p, \ast\right)$}

\subsection{The state-space} In the course of perceiving a physical item, the corresponding sensory input invokes an activity pattern in the neuronal field it is projected to. That way, each physical item can be represented by a binary pattern, in which $1's$ indicate active neurons, while $0's$ indicate inactive ones. Due to the size and structural complexity of the neuronal correlate, patterns are described by high-dimensional random binary vectors. These patterns are the states of the cognitive system. The state-space therefore is
\begin{equation}\label{eq:state-space}
\mathbb{X} = \left( \mathbb{X}^N_q, d\right). 
\end{equation}
$N>100$ is its dimension, $q$ is the mean activity of a state, and $d$ is some metric on $\mathbb{X}^N_q$. 


The state-space is a (metric) Hamming space allowing for some similarity measure derived from the distance $d$. This measure should respect both:
the closeness of two states as well as their distinctiveness as points in the state-space. A cosine-similarity only reveals information about closeness since it is locally defined. In a probabilistic setting, two points are the more difficult to distinguish, the less likely it is to find another state at random which is 'in between' the two. To capture this, the definition of similarity must contain global information about the state space. 

\begin{defi}[Similarity]
The similarity of two states having distance $d$ from each other is
\begin{equation}\label{eq:similarity}
S(d) := e^{-\kappa F_\mathbb{X} (d)},  \qquad \kappa>1
\end{equation}
where $F_\mathbb{X}(d) = \mathbb{P}_\mathbb{X}[D \leq d]$ is the distribution function for distances on $\mathbb{X}$.
\end{defi} 

Different items are represented by uncorrelated states, while similar items will be represented by similar states. $\kappa > 1$ is chosen to have highest sensitivity with respect to almost identical or near-by states. 

\subsection{The operations}
The two operations to be defined on the state space correspond to binding and bundling. Two items are (associatively) bound to each other, if one can be retrieved by cueing with the other item. The corresponding formal operation is {\it multiplication} $\ast$, which is defined in eq \ref{eq:operations}. Binding of items happens by simultaneously activated components in the neural pattern. This similarity measure directly relates to a recall probability or accuracy of retrieval.
 
Bundling means collecting items by adding their respective states. 
Assume that two neurons $X$ and $Y$ converge on a third neuron $Z$. If both are inactive, i.e. $x=y=0$, neuron $Z$ will also be, $z=0$, while if both are active, $Z$ will be active, i.e. $1+1 = 1$. If only $X$ or $Y$ is active, it depends on some threshold, whether $Z$ is active. If the activation threshold is low, $p \approx 0$, $Z$ is likely to be active, while if if the activation threshold is high, $p\approx 1$, $Z$ will remain inactive. Addition $x_p$ is defined in eq \ref{eq:operations}. 

\begin{equation}\label{eq:operations}
\begin{array}{c|cc}
	{\ast} & 0 & 1\\ \hline
	0& 1 & 0\\
	1& 0  & 1
\end{array}
	\qquad \qquad 
\begin{array}{c|cc}
	{+_p} & 0 & 1\\ \hline
	0& 0& \zeta\\
	1& \zeta & 1
\end{array}
\end{equation}
where $\zeta \in \{0,+1\}$ is random with ${\mathbb P}[\zeta=0] = p$. This completes the definition of the algebra used to calculating with cognitive states.
\begin{equation}\label{eq:algebra}
\Big( \mathbb{X},+_p,\ast \Big) 
\end{equation} 
In the following, its elementary properties are further investigated. What properties are already implied by this elementary algebra and how much of empirical findings can be already described by those?

\paragraph{Bundling preserves sequential information in the memory list}

Usually, bundling is realised by vector-addition \cite{schlegel2020comparison}, which is commutative and associative, so that $x+(y+z) = (x+y)+z = z + (y+z)$, i.e. the order of components doesn't matter. That is: If addition is associative, sequential order information is lost! 

\begin{obs}
For $0 < p < 1$, addition $+_p$ is not associative.
\begin{equation}\label{eq:non-asso}
x+_p(y+_pz) \not= z +_p (y+_p x)
\end{equation}
\end{obs}
Note that, if $p=1$, addition equals component-wise $AND$, while for $p=0$, addition is component-wise $OR$. These operations are associative.

In the following, the state resulting from left-associative addition is denoted by ${\bf L}$, i.e. ${\bf L} = (x+_p y) + z$, while the state resulting from right-associative addition is denoted by ${\bf R} = x +_p (y+_p z)$.
For the sake for readability, I will write $+ = +_p$ in the following, while assuming that $p = \frac{1}{2}$. 

\begin{figure}[h]
{\center
\includegraphics[height=1.8in]{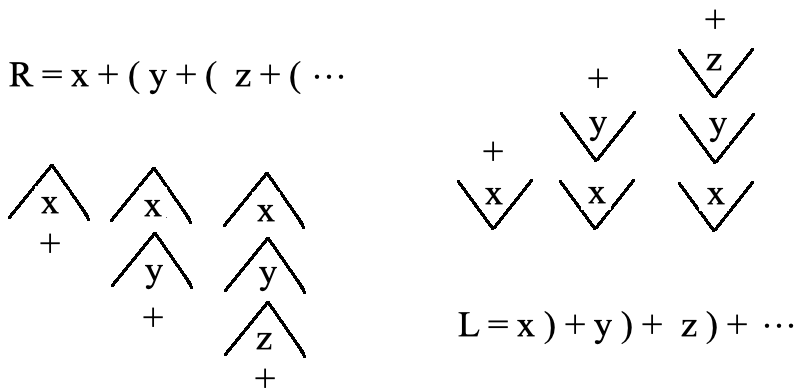} 
\caption{In right-associative $\bf R$ addition, early items are kept prominent, while in left-associative $\bf L$ addition, later items are superposed on earlier ones.}\label{fig:StorageModes}
}
\end{figure}

\subsection{The states representing a memory list}
$\bf L$ and $\bf R$ states can be constructed for a list of any length. Construction starts from a pre-experimental state $\eta$ and proceeds by iteratively adding items to the memory states $\bf L$ and $\bf R$ according to left-associative addition and right-associative addition to the respective branch as follows: For the $\bf L$-state
	\begin{eqnarray*}\label{eq:LR}
	{\bf L}^0&=&   \eta \\
	{\bf L}^a &=&  \eta+a)\\
	{\bf L}^b &=&  \eta+a)+ b)  \\
	&\vdots&\\
	{\bf L}^\Lambda &=& \Big( \big( ((\eta+a) + b ) + c )+\hdots )+f\big) + g\Big),
	\end{eqnarray*}
	while for the $\bf R$-state
	\begin{eqnarray*}
	{\bf R}^0&=&  \eta \\
	{\bf R}^a &=&  \eta + ( a \\
	{\bf R}^b &=&  \eta + (a + ( b  \\
	&\vdots&\\
	{\bf R}^\Lambda &=& \Big( \eta + \big(a +  (b + (c +(\hdots+(f+g) ) ) \big) \Big)
	\end{eqnarray*}

After its sequential presentation, the memory list $\Lambda = (A,B,C,\hdots)$ is thus represented by the two states
\begin{eqnarray}
{\bf L} &=&  \Big( \Big( ((\eta+a) + b ) + c \big)+d )+f) + g\Big) \\
{\bf R} &=&  \Big( \eta + \big(a +  (b + (c +(d+(f+g) ) ) \big) \Big), 
\end{eqnarray}

\begin{figure}[h]
{\center
\includegraphics[height=3in]{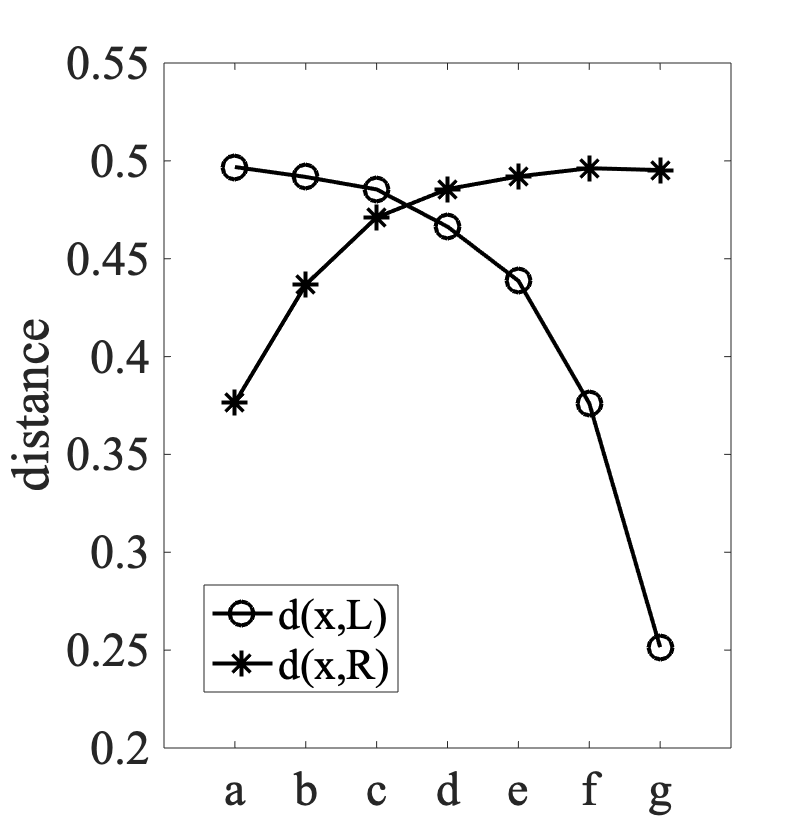} 
\caption{{\bf Distance profiles} of the two states $\bf L$ and $\bf R$ as in eq \ref{eq:distancegradients} ff. $\bf L$ has smallest distances to the most recent items, while $\bf R$ is closest to the early list items. }\label{fig:DistancesGradients}}
\end{figure}

In \cite{murdock1982theory} $\eta$ is assumed to be empty, while in \cite{franklin2015memory} it comprises a holographic collection of items and item-item associations. $a,b,c,\hdots$ are the cognitive states representing the physical list items $A,B,C,\hdots$. These states preserve the serial order of items in the memory list in that distances change monotonously along subsequent items, see Fig. \ref{fig:DistancesGradients}
\begin{eqnarray}\label{eq:distancegradients}
	d(\eta,{\bf R}) &<& d(\eta, {\bf L})\\
	d(a,{\bf L}) &>& d(b,{\bf L}) > d(c,{\bf L}) >\hdots \\
	d(a,{\bf R}) &<& d(b,{\bf R}) < d(c,{\bf R}) <\hdots  
\end{eqnarray}
Correspondingly, both states inherit serial order in that item distances increase along $\bf R$, while they decrease along $\bf L$, see Fig \ref{fig:DistancesGradients}. These distance gradients directly translate into activation gradients.

\subsection{Implied activity gradients}
From the concept of similarity, two other concepts can be immediately derived: activation and memory strength. The intuition is closely related to the idea of a projection. Given that the memory state $\bf M$ represents a memorised list, and that a cue item is presented. The cue item activates the memory state more, the more similar it is to that memory state \cite{hintzman1984minerva}. Conversely, the more the corresponding memory element is engraved in the memory state, the more the memory state is activated by the cue state.    

\begin{defi}[Activation]
Let $\bf M$ be a memory state constructed during representing some memory list. 
A cue state $x$ activates the memory state $\bf M$ according to their similarity, see eq \ref{eq:similarity}
\begin{equation}\label{eq:activation}
\alpha_{\bf M}(x) := S\big(d(x,{\bf M})\big).
\end{equation}
The activation gradient of $\bf M$ is the vector $\alpha_{\bf M}$ with components $\alpha_{\bf M}(x)$, where $x$ is a state representing a list item.  
\end{defi}

\begin{figure}[h]
{\center
\includegraphics[height=3in]{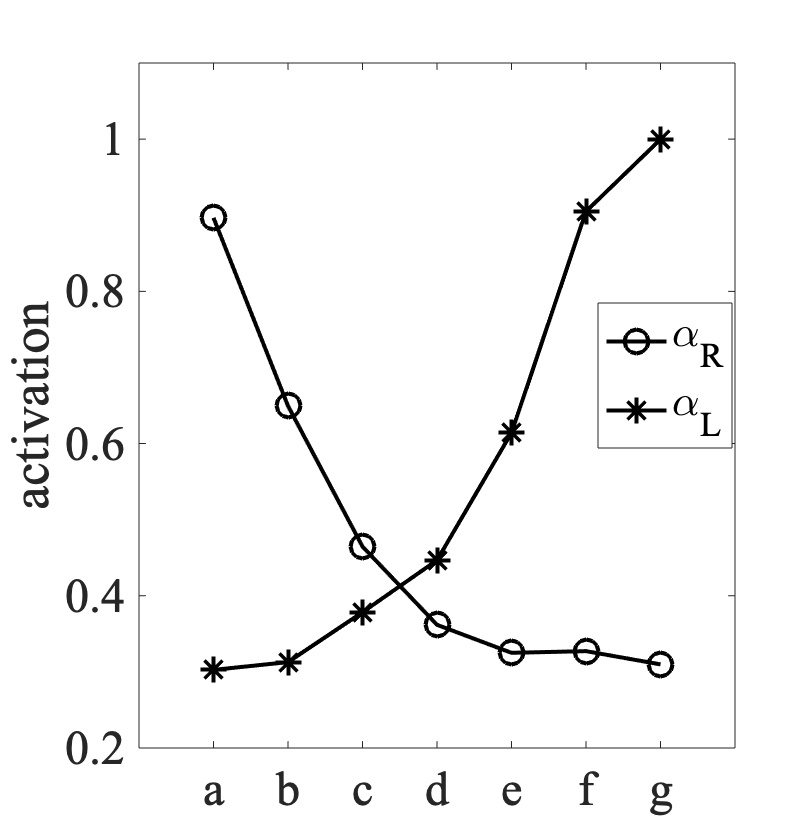}
\caption{{\bf Primacy and recency gradients} implied by the two states $\bf L$ and $\bf R$ are $\alpha_{\bf R}$ and $\alpha_{\bf L}$. }\label{fig:ActivationGradients}
}
\end{figure} 

In terms of strength theory, $\alpha_{\bf M}(x)$ is the strength by which $x$ is memorised in $\bf M$. One might also call $\alpha_{\bf M}(x)$ the familiarity of $x$ given $\bf M$.

Consequently, the distance gradients in eq \ref{eq:distancegradients} ff directly translate into activity gradients, see Fig. \ref{fig:ActivationGradients}. Since activation as well as strength are increasing functions of similarity and hence decreasing functions of distance,  ${\bf L}$ implies a {\it recency gradient} $\alpha_{\bf L}$ , while ${\bf R}$ implies a {\it primacy gradient} $\alpha_{\bf R}$. 
%

Activation gradients are nowadays widely accepted to play an important role in working memory. Various mechanisms have been discussed as sources of these gradients, see
\cite{oberauer2003understanding}. In many models including TODAM, TCM and SOB, these gradients are separately modelled and superimposed on top of the model. In contrast, these gradients directly result from the bundling operation defined in eq \ref{eq:algebra} and its non-associativity: While non-associativity preserves information about serial order, right-associative addition and left-associative addition imply the primacy and the recency gradient, respectively. 
%

%
%
%

\subsection{The response function for recognition and recall}
After presentation of a memory list, the participant has to fulfil some task. Most cognitive tasks involve cues such as cued item recognition or cued recall, associative or serial. The answer the participant gives is the result of a decision process which depends on both, the memory state as well as the cue. The response function in recognition only depends on familiarity, while the response function in recall additionally depends on distinctiveness \cite{murdock1982theory}. Thus it is reasonable to make the response function a function of activation as defined in eq \ref{eq:activation}.

\begin{defi}[Response function]
The response function given a cue $x$ facing the memory state $\bf M$ is an increasing function of induced similarity, e.g.
\begin{equation}
\Phi(x\: | \:{\bf M}) = \alpha_{\bf M}(x)
\end{equation}
\end{defi}

Accordingly an activation gradient directly translates into a serial position curve. Particularly, the recency effect refers to the activation gradient of the $\bf L$-state, while the primacy effect corresponds to the activation gradient of the $\bf R$- state.

\begin{figure}[h]
{\center
\includegraphics[height=2.2in]{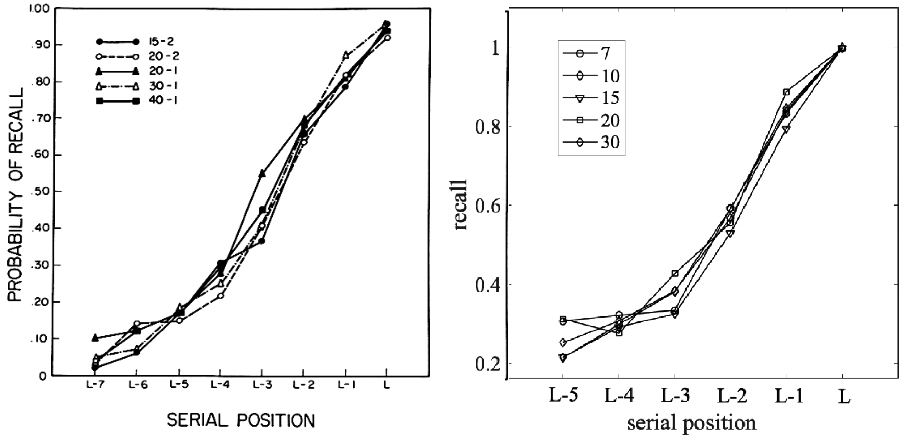}
\caption{{The recency effect does not depend on list length.} Left: Experimental data from Murdock, Right: Simulated data from the model for various list length'. } \label{fig:recencyeffect}
}
\end{figure}

Experimental data indicate that the recency effect does not depend on list length and shows a slightly sigmoid curve shape, see Fig \ref{fig:recencyeffect} (left). Both empirical observations are well captured by the modelling algebra proposed, see Fig \ref{fig:recencyeffect} (right). 
 
\section{Application to some basic cognitive tasks}
In this section some examples are presented to demonstrate how the formalism works, i.e. how to describe tasks such as cued recall in this formalism. Results are direct consequences of the algebra defined, i.e. no further assumptions are made. 
In the following only the $\bf R$-state is concerned, i.e. states are bundled according to right-associative addition, while corresponding brackets are skipped for the sake of readability.

\subsection{Repetition increases strength}
\begin{figure}[h]
{\center
\includegraphics[height=2.5in]{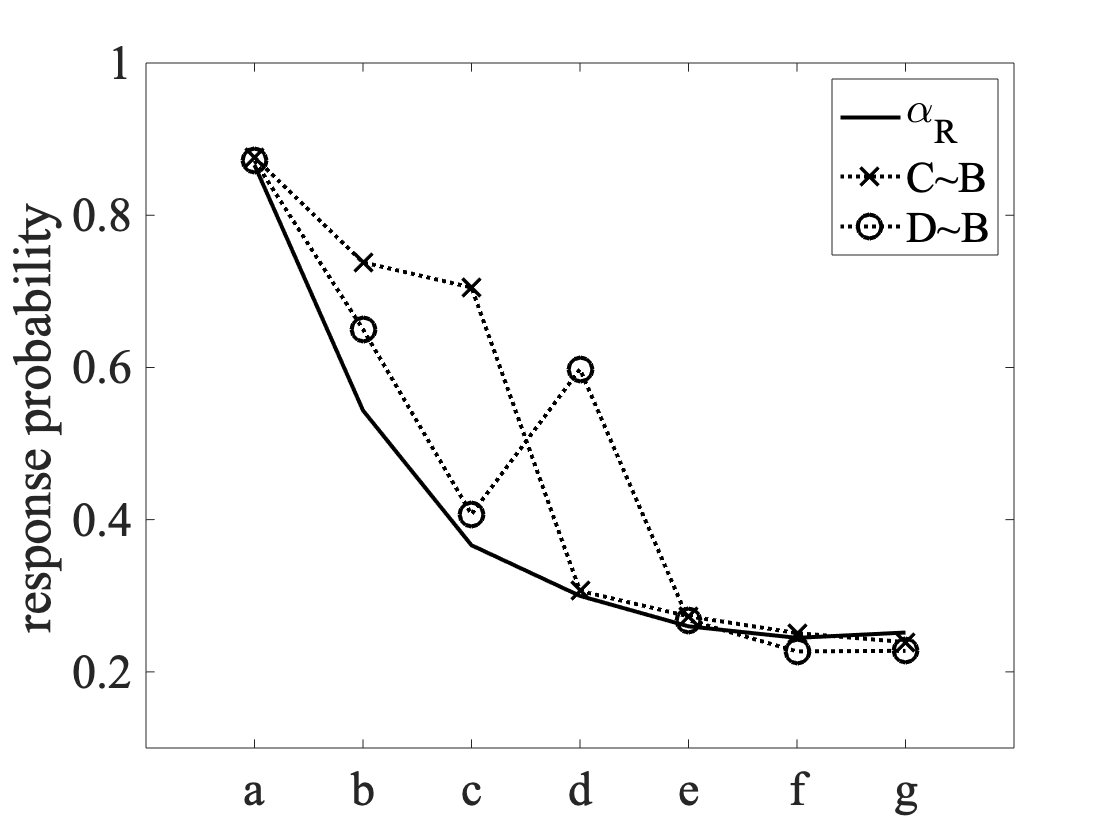}
\caption{{\bf The effect of similar items on activation} The solid black curve is the activation profile of state $\bf R$ for the list $\Lambda$ in which all items are different. Doted lines are the profiles if that list contains one item, e.g. $C$ or $D$, which is similar to item $B$.}\label{fig:similarItems}
}
\end{figure}

It is intuitively expected that a repeated occurrence of an element in a list will increase its coding strength. This effect is indeed observed in the model. As a benchmark, consider the list $\Lambda = (A,B,C,D,\hdots)$, in which all items are different. In $\Lambda^{(1)}$ a neighbouring pair is similar, e.g., $B \sim C$. In $\Lambda^{(2)}$, $B\sim D$ and so forth. $k$ can be regarded as the lag from $B$ until the similar item. Fig \ref{fig:similarItems} shows the serial position curves for lists $\Lambda$, $\Lambda^{(1)}$, and $\Lambda^{(2)}$. Note that the coding strength of $B$ is increased by any other item which is similar to $B$, while the strengthening is greater, the smaller the lag is, i.e. the effect of $C\sim B$ on the coding strength of $B$ is larger than the effect of $D\sim B$. 

\subsection{Cued recall}
\subsubsection{Cued associative recall}
In this task, the participant is presented a paired memory list $\big( A-X,  B-Y, C-Z, \hdots \Big)$. After memorizing this list, a memory item, i.e. a member of some pair, is presented as a cue, and the participant is asked to identify the memory item, which was bound to that cue item. The memory state corresponding to the paired list is
\begin{equation}
{\bf R} = \eta + a*x + b*y + c*z + \hdots,  
\end{equation}
where $a*x$ is the state representing the binding between items $X$ and $A$ in the list.
 
\begin{figure}[h]
{\center
\includegraphics[height=3in]{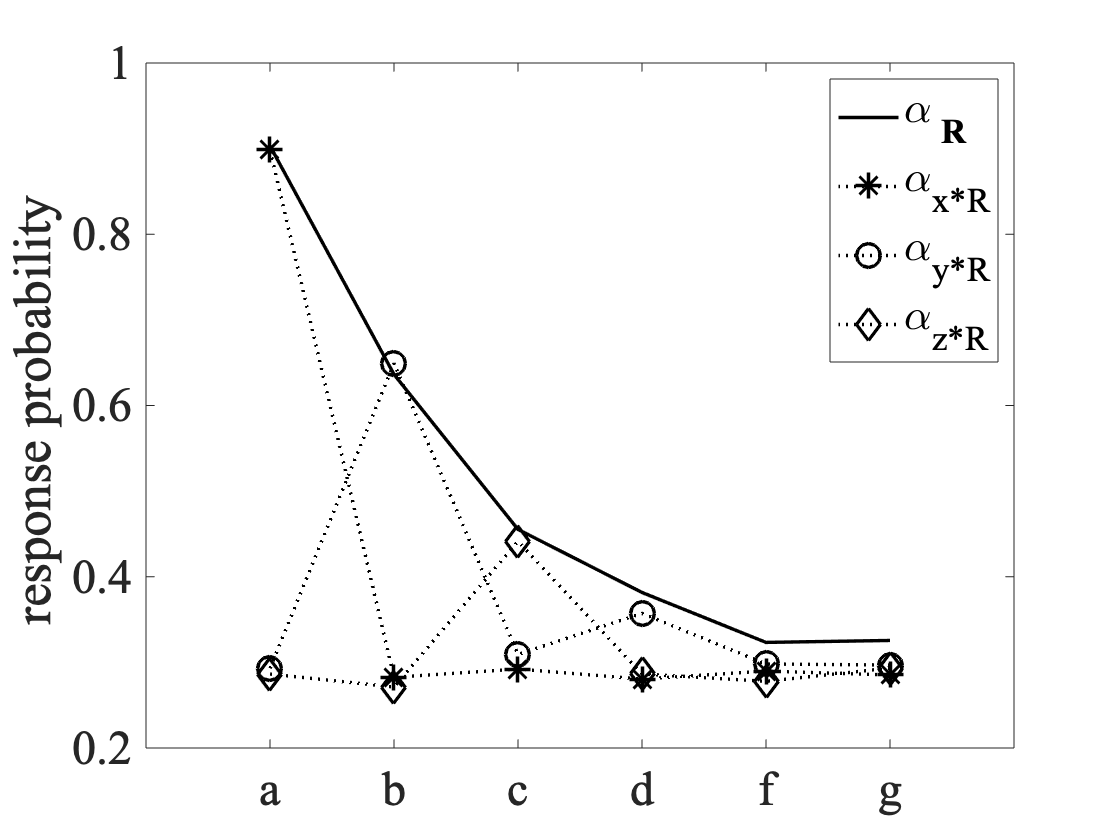}
\caption{{\bf Cued Recall}: Given cues such as $x$, $y$, $z$, the corresponding activations $\alpha_{x*\bf R}, \alpha_{y*\bf R}, \alpha_{z*\bf R}$ are plotted, see eq \ref{eq:xR}. The cue $x$ causes the activation $\alpha_{x*\bf R}$ to have a peak at the corresponding item, which is $a$. $\alpha_{\bf R}$ is the activation profile of the ${\bf R}$-state.}\label{fig:Retrieval2}
}
\end{figure} 

When a memory item $X$ is presented as a cue and the task is to retrieve the item which is bound to $X$ in the list, consider the activation of 
\begin{equation}\label{eq:xR}
x*{\bf R} = x*\eta + \: a +\:  a*b*y + a*c*z + \hdots.
\end{equation}
The activation $\alpha_{x*{\bf R}}$ attains its maximal value for $\alpha_{x*{\bf R}}(a)$, see Fig. \ref{fig:Retrieval2}. Thus the cue $X$ activates the $A$ component most, so that the participant will answer {\it " $X$ is bound to $A$."} , with some probability. Analogously, if the cue is $Y$, the activation $\alpha_{y*{\bf R}}$ attains it maximum in $B$, so that $B$ is retrieved, and so forth. These maximal points form a curve, which is identical to the activation gradient $\alpha_{\bf R}$.

\subsubsection{Retrieval from similar contexts}
\begin{figure}[h]
{\center
\includegraphics[height=3in]{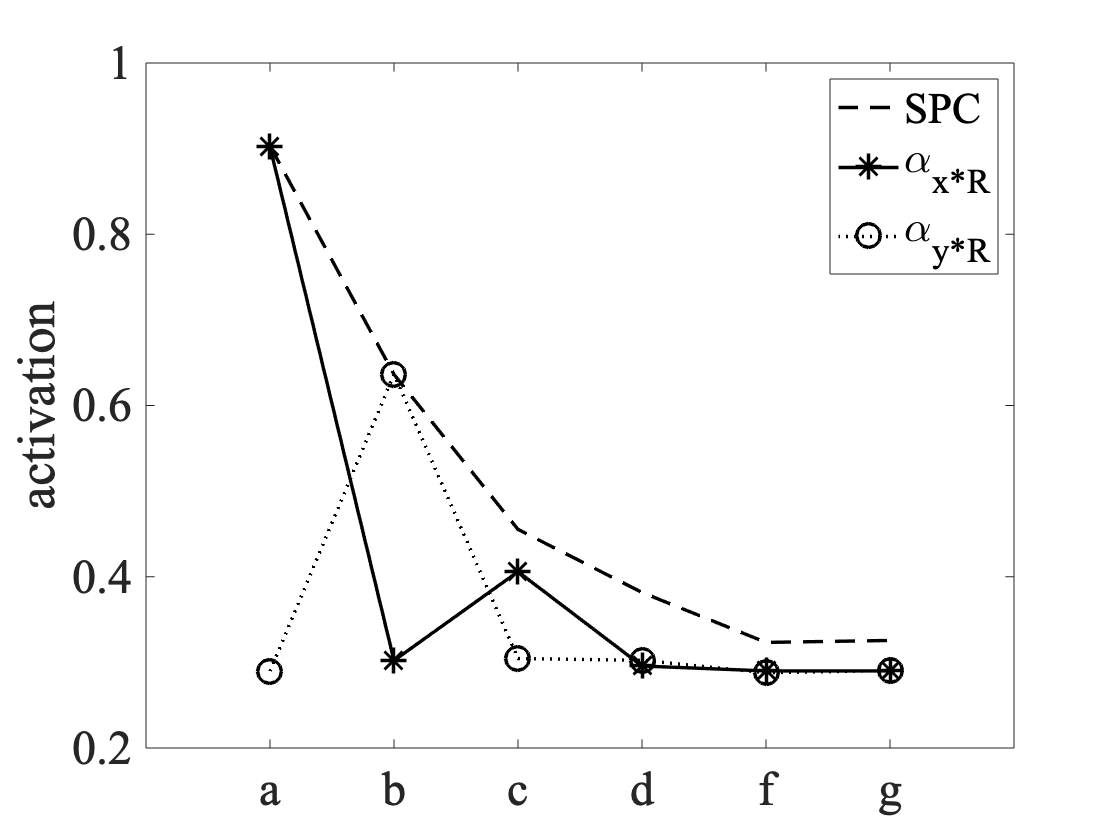}
\caption{{\bf Recall of items}: Activation profiles of $\alpha_{x*\bf R}$ ($\ast$), and $\alpha_{y*\bf R}$ ($\circ$). The activation profile $\alpha_{\bf R}$ of the ${\bf R}$ state by distinct list items is shown as a reference.}\label{fig:AssociativeContext}
}
\end{figure}

Assume that the paired list $\big( A-X,  B-Y, C-\tilde{X}, D-Z, \hdots \big)$ is given, in which items $A$ and $C$ are bound to similar contexts $X$ and $\tilde{X}$. The corresponding state yields
\begin{equation}\label{eq:similar}
{\bf R} = \eta + a*x + b*y + c*\tilde{x} + \hdots. 
\end{equation}
Cueing with $X$ will not only retrieve $A$ but also $C$, just to a lesser extend. The effect of cueing with $x$ is displayed when considering the activation gradient $\alpha_{x*{\bf R}}$, see Fig. \ref{fig:AssociativeContext}. The gradient has two peaks, one at $a$ and a weaker one at $c$, saying that cueing with $X$ reveals two items, $A$ and $C$. Cueing with $Y$ uncovers only one, which is $B$. 

In the recall task, the participant has to make a choice between the two alternative items bound to $X$. Thus invoking Luce's choice axiom, the probability to recall $X$ yields
\begin{equation}
P(a|x) = \frac{\alpha_{x*\bf R}(a)}{\alpha_{x*\bf R}(a)+\alpha_{x*\bf R}(c)} 
\end{equation}
which is less than the probability to recall $a$ without an alternative. The existence of an item similar to the cue impairs the corresponding recall.


\subsection{Putting things together: The serial position curve}

\begin{figure}[h]
{\center
\includegraphics[height=3in]{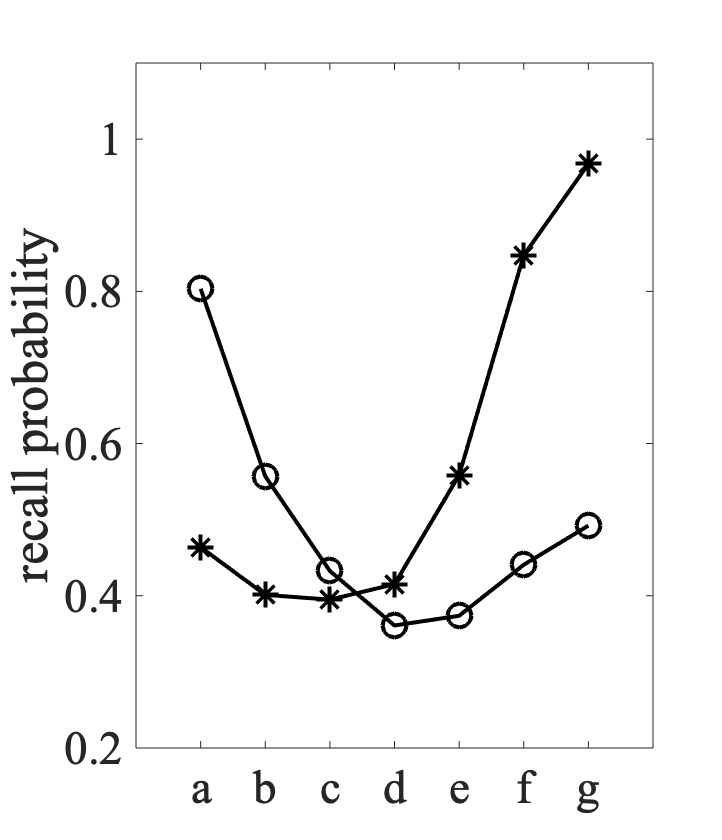}
\caption{{\bf The Serial Position Curve} is simulated for different pairs of parameters according to eq. \ref{eq:SPC}. It shows a strong recency effect and a weak primacy effect for $\rho = 0.4, \lambda = 0.9$, while for $\rho = 0.9, \lambda = 0.4$ there is a strong primacy effect and a weak recency effect.}\label{fig:SPC}
}
\end{figure}

During memorizing a list, the two states $\bf R$ and $\bf L$ are constructed. Since there is no a priori reason to favour one over the other, I assume that both cognitive states $\bf L$ and $\bf R$ coexist and are the components of a {\it memory state} {\bf M}, 
\begin{equation}\label{eq:S}
{\bf M} =\begin{pmatrix} {\bf L }\\ {\bf R}\end{pmatrix}.
\end{equation}
A single cue thus activates both components. The total activation of the memory state $\bf M$ is a linear combination of the activation gradients of its two components.
\begin{equation}\label{eq:SPC}
\alpha_{\bf M} = \rho \: {\bf R} + \ell \: {\bf L}
\end{equation}
where $\rho$ and $\ell$ are non-negative parameters governing the mixture of respective activations. 
The response function to a cue is $\Phi(x|\alpha_{\bf M})$, so that the serial position curve is the graph $\Phi(x|\alpha_{\bf M})$, where $x$ is a state representing a list item, see Fig. \ref{fig:SPC}. 

The serial position curve thus results from the linear combination of the primacy gradient $\alpha_{\bf R}$ and the recency gradient $\alpha_{\bf L}$.  As seen in Fig \ref{fig:SPC}, a large $\rho$ together with a small $\ell$ makes the recency effect, while a small $\rho$ together with a large $\ell$ leads to a prominent primacy effect. The relative strength of the primacy and the recency effect will generally depend on the experimental set-up, including the task to be performed.  For example in recognition and in cued (probed) recall the serial position curve shows a strong recency effect, while the primacy effect is weak. Strong primacy effects are seen in forward recall, while recency effects are strong in backward or in free recall.

\section{Conclusion and out-look}
In the previous sections, an elementary algebra ( eq. \ref{eq:algebra} ) for storage and retrieval of information in basic cognitive tasks was proposed. The aim was not to present a full-blown model but to investigate how far one can get with the algebra alone. 

Item information and associative information are represented by two operations, bundling and binding, respectively. If bundling is realised by an associative operation such as ordinary (vector-) addition, information about sequential order is lost. On the other hand, tasks such as serial recall require that order information. Consequently in corresponding models order information has be has to be implemented separately. This can be achieved by postulating serial position markers, chaining by associative mechanisms between consecutive items, or weight functions varying over serial position governing the recency and the primacy effect. 

This is different in the approach presented: Information about sequential ordering is preserved. This is due to the non-associativity of the addition operation by which item information is bundled into a memory state. Reading from that state thus reveals order information necessary to related tasks, which is represented by corresponding gradients. Activation gradients are implied rather than postulated separately. The serial position curve comes as a linear combination of both. Its shape is concave-up and asymmetric as observed as a typical experimental finding, see Fig \ref{fig:expCurve} for experimental data and Fig \ref{fig:SPC}  for simulations of our model.

As already mentioned, the aim was not to present a full-blown model but to investigate how far one can get with the algebra alone. So it does not come as a surprise that several experimental observations were not captured. For example, while the recency effect does not depend on list length, the primacy effect does. This robust finding cannot be explained by our algebra alone but needs an additional assumption about attention, which then imposes an additional constraint on den attention gradient. Furthermore, serial recall can not be described by our algebra alone but needs an additional assumption such as output-suppression, as supposed in many models, or an other feedback mechanism, see \cite{franklin2015memory}. 


The cognitive algebra proposed appears to provide a reasonable basis for modelling since it generically implies several features that fit empirical observations quite well, in a qualitative sense in that no attempt was made to fit data. Modelling then could consist in carefully adding assumptions on top of the cognitive algebra such as discussed above.   


\end{document}